\documentclass[11pt]{article}

\usepackage{subfig}
\usepackage{graphicx}
\usepackage{multirow}

\usepackage[numbers]{natbib}

\usepackage{color}
\input{amssym.def}

\title{Least-perimeter partition of the disc into $N$ regions of two different areas}
\author{F.~J. Headley and S.~J. Cox \\ 
Department of Mathematics, Aberystwyth University, SY23 3BZ, UK.}
\date{December 2018}

\begin{document}

\maketitle

\begin{abstract}
We present conjectured candidates for the least perimeter partition of a disc into $N \le 10$ regions which 
take one of two possible areas. We assume that the optimal partition is connected, and therefore enumerate 
all three-connected simple cubic graphs for each $N$. Candidate structures are obtained by assigning 
different areas to the regions: for even $N$ there are $N/2$ regions of one area and $N/2$ regions of the 
other, and for odd $N$ we consider both cases, i.e. where the extra region takes either the larger or the 
smaller area. The perimeter of each candidate is found numerically for a few representative area ratios, and 
then the data is interpolated to give the conjectured least perimeter candidate for all possible area ratios. 
At larger $N$ we find that these candidates are best for a more limited range of the area ratio.
\end{abstract}

\section{Introduction}

Due to their structural stability and low material cost, 
energy-minimizing structures have a wide array of 
applications~\cite{mousse13}. In engineering an example is the Beijing 
Aquatics Centre, which uses slices of the Weaire-Phelan 
structure~\cite{WeaireP94b} to create a lightweight and strong but 
beautiful piece of architecture.

The Weaire-Phelan structure is a solution to the celebrated Kelvin 
problem, which seeks the minimum surface area partition of space into cells 
of equal volume~\cite{kelvin87}. This builds upon the well-known 
isoperimetric problem, concerning the least perimeter shape enclosing a 
given area~\cite{morgan4th}. Extending this idea to many regions with 
equal areas has led to further rigorous results for optimal structures,
for example the proof of the honeycomb conjecture~\cite{hales01}, 
the optimality of the standard triple bubble in the plane~\cite{wichiram04} 
and of the tetrahedral partition of the surface of the 
sphere into four regions~\cite{engelstein09}.

If the areas of the regions are allowed to be unequal, then the problem 
of seeking the configuration of least perimeter is more difficult. For 
$N=2$ regions in ${\Bbb R}^3$, the double bubble conjecture has been 
proved~\cite{hutchingsmrr02}, and, in the plane, the extension of the 
honeycomb to two different areas (bidisperse) has led to conjectured 
solutions~\cite{fortest01b}. There has also been some experimental work 
that sought to correlate the frequency with which different configurations of 
bidisperse bubble clusters (which, to a good approximation, minimize 
their surface area~\cite{mousse13}) were found with the least perimeter 
configuration~\cite{vazca04}.
   
Minimal perimeter partitions of domains with a fixed boundary have also 
generated interest, for example a proof of the optimal partition of the 
disc into $N=3$ regions of given areas~\cite{caneter04}, and many 
numerical conjectures, 
e.g.~\cite{tomonaga74,bleicher87,Coxf10,bogoselo16}. Such results may 
lead to further aesthetically pleasing structures like the Water Cube 
but that are truly foam-like, including their boundary, rather than 
being unphysical sections through a physical object.

In this work we seek to generate and test, in a systematic way, 
candidate partitions of domains with fixed boundary. Due to the complexity, and in particular the large 
number of candidates, we restrict ourselves to a two-dimensional (2D)
problem. Thus, we enumerate all partitions of a disc and evaluate the 
perimeter of each one to determine the optimal configuration of the 
regions.
  
 \begin{figure}
 		\centering
 		\subfloat[$P=$6.304 \quad\quad\qquad \quad\quad \quad (b) $P = 6.272$]{\includegraphics[width=0.5\linewidth]{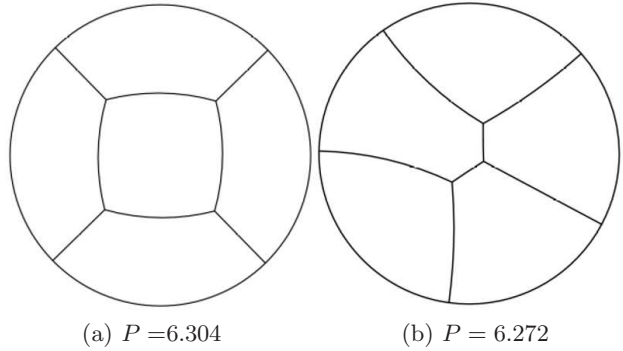}}
 		\caption{The two different partitions of the disc into $N=5$ regions of equal area. The structure on the right has least perimeter $P$.}
 		\label{fig:two_5}
 \end{figure}

As the number of regions $N$ increases then so does the complexity of 
the system and for $N \ge 5$ numerical methods must be employed. For 
example, figure \ref{fig:two_5} shows the two three-connected ``simple" 
partitions of the disc into $N=5$ regions with equal area. The 
difference in perimeter comes from the different structural arrangements 
of the arcs separating the regions. If we allow three regions to have 
one area and the other two a different area then there are 20 possible 
structures. When $N = 10$ this number increases to 314,748.

We will use combinatorial arguments to enumerate the graphs corresponding to 
all possible structures.  We recognise that all structures must obey 
Plateau's laws~\cite{plateau73}, a consequence of perimeter 
minimization~\cite{taylor76}, which state that edges have constant 
curvature and meet in threes at an angle of $2\pi/3$. Rather than 
applying these directly, we will rely on standard numerical minimization 
software to determine the equilibrated configuration for each choice of 
$N$ and areas.

\section{Enumeration and evaluation of candidate structures}

\begin{figure}
\centering
\includegraphics[scale=0.35]{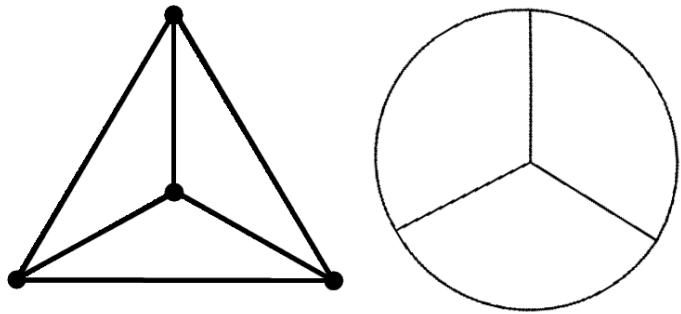}
\caption{A simple cubic three-connected planar graph with three regions of equal area, and its associated minimal-perimeter monodisperse partition of the disc.} 
\label{fig:graph}
\end{figure}

As the basis for enumerating possible partitions of the disc, we 
consider each candidate structure as a simple, three-regular (cubic), 
three-connected planar graph (figure \ref{fig:graph}). There is a one-to-one 
correspondence between these graphs and the candidate solutions to the 
least perimeter partition.

\begin{figure}
\center
\includegraphics[scale=0.25]{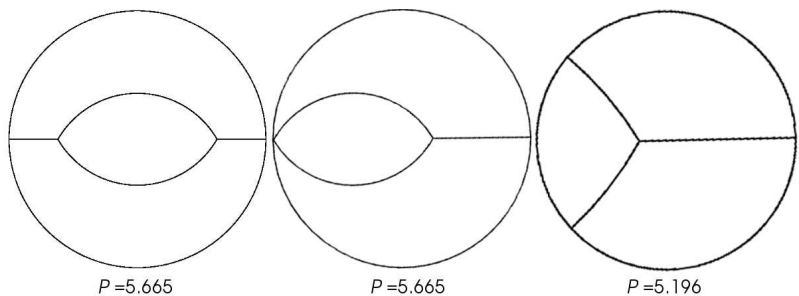}
\caption{The central two-sided region in this non-simple, two-connected structure (left) can be moved so that one of its vertices touches the boundary (middle) without changing the perimeter $P$ of this configuration. Once there, a change in topology results in a drop in the perimeter and a simple, three-connected, state (right).} 
  \label{fig:notsimple}
 \end{figure}

The assumption of planarity is natural, since these graphs must be 
embeddable in the 2D disc. The assumption that the graphs 
are three-regular follows from Plateau's laws. We assume that the graphs 
are simple and three-connected because any two edges sharing two vertices 
can be decomposed into a configuration with lower perimeter $P$. An example 
is shown in figure \ref{fig:notsimple}: moving the lens-shaped 
region to the edge of the disc results in a change in topology and a 
reduction in perimeter. A similar reduction in perimeter can be achieved 
in structures with more regions by moving a lens towards a threefold vertex 
and performing the same change in topology.

We use the graph-enumeration software CaGe~\cite{brinkmanncage} to generate every graph and an associated embedding for each value of $N$. This information is stored as a list of vertices, each with an $(x,y)$ position and a list of neighbours. The number of graphs for each $N$ is given in Table \ref{tab:numcands}.

The Surface Evolver~\cite{brakke92} is finite element software for the minimization of energy subject to constraints.
We convert the CaGe output into a 2D Surface Evolver input file~\cite{Coxf10}, in which each edge is represented as an arc of a circle and the relevant energy is the sum of edge lengths. The cluster is confined within a circular constraint with unit area, and we set a target area for each region. The Evolver's minimization routines are then used to find a minimum of the perimeter for each topology and target areas.

If an edge shrinks to zero length during the minimization, this is not a topology that will give rise to a stable candidate, since four-fold vertices are not minimizing. We therefore allow topological changes when an edge shrinks below a critical value $l_c$ (we use $l_c = 0.01$, which is less than 1/50th of the disc radius). This prevents time-consuming calculation of non-optimal candidates, but does result in some solutions being found repeatedly as the result of different topological changes on different candidates.

Our aim is to consider bidisperse structures, in which each region can take one of two possible areas. We define the area ratio $A_r$ to be the ratio of the area of the large regions to the area of the small regions, so that $A_r > 1$. When the smaller regions are very small, the precise area ratio changes the total energy only very little, so we consider $A_r$ up to 10. (The highest area ratio at which we find a change in the topology of the optimal structure is $A_r = 8.35$.)

\begin{figure}
\center
{\includegraphics[width=0.7\textwidth]{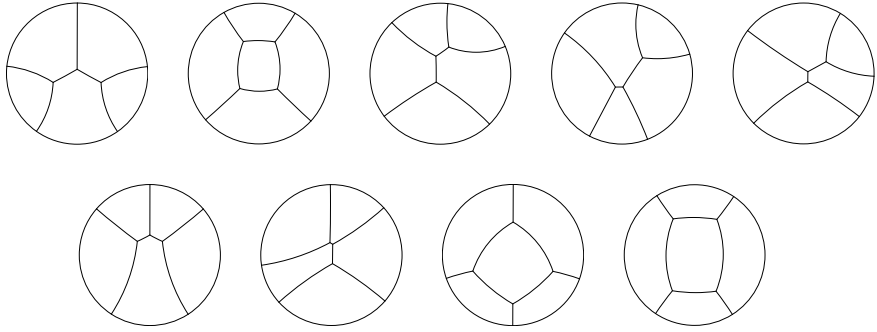}}
\caption{
All partitions of the disc into $N=5$ regions with area ratio $A_r = 2$ and three large and two small regions ($5_{32}$). The candidates are shown in order of increasing perimeter (left to right, top row then bottom row). Note how the motif of the two structures in figure~\protect\ref{fig:two_5} is repeated with different arrangements of the two possible areas.
}
\label{fig:bidis}
\end{figure}

To reduce the number of possible candidates, we stipulate that the number of regions of each area are equal (when $N$ is even) or (when $N$ is odd) as close as possible. In the latter case, we consider both possibilities: one extra large region or one extra small region; see figure \ref{fig:bidis}. We label a configuration with $N_L$ large regions and $S= N-N_L$ small regions as $N_{LS}$. For each graph we permute all possible arrangements of the areas of the $N$ regions (with some redundancy). 

For example, for $N=3$, there is only one possible graph (figure \ref{fig:graph}), in which three lines meet together in an internal vertex, as for the monodisperse case. Since $N$ is odd we consider $3_{21}$ and $3_{12}$ separately. In the first case there are three possible permutations of the areas assigned to the three regions, but all three are clearly equivalent through a rotation, so there is only one candidate for which the perimeter must be evaluated. In the second case there are also three possible permutations of the area, but again only one candidate needs to be minimized.

The number of graphs and the number of area permutations rises rapidly. We therefore treat only values of $N$ between 4 and 10. The number of candidates that we evaluate and the number of structures that are actually realized is shown in Table \ref{tab:numcands}.

  \begin{table}
 \centering
\begin{tabular}{|r|r|r|r||r|r|r|r|r|}
 \hline
 $N$ & Graphs & Permutations & Total Foams & $A_r = 2$ & $A_r = 4$ & $A_r = 6$ & $A_r = 8$ & $A_r = 10$ \\ \hline
4       & 1      & 4            & 4        &  4 & 4 & 4 & 3 & 3     \\ \hline
\multirow{2}{*}{5}    & \multirow{2}{*}{2}     & \multirow{2}{*}{10}  & \multirow{2}{*}{20}   & 9 & 7 & 6 & 5 & 6   \\ \cline{5-9}
				& & &	   & 9 & 8 & 7 & 8 & 8 \\ \hline
6       & 5      & 20           & 100       & 31 & 25 & 19 & 17 & 19  \\ \hline
\multirow{2}{*}{7}  & \multirow{2}{*}{14} & \multirow{2}{*}{35}  & \multirow{2}{*}{490} & 136 & 100 & 74 & 76 & 76 \\ \cline{5-9}
                                & & &       & 139 & 96 & 78 & 75 & 76 \\ \hline
8       & 50     & 70           & 3500      & 711 & 495 & 377 & 358 & 380  \\ \hline
\multirow{2}{*}{9} & \multirow{2}{*}{233} & \multirow{2}{*}{126}  & \multirow{2}{*}{29358} 
						& 3716 & 2619 & 2072 & 1949 & 1962      \\ \cline{5-9}
				& & &       & 3608 & 2562 & 2074 & 1958 & 1971 \\ \hline
10      & 1249   & 252          & 314748    & 22145 & 15217 & 12536 & 11990 & 12008  \\
\hline  
\end{tabular}
\caption{For each number of regions $N$ we show the number of simple, cubic, three-connected graphs, the number of permutations of the two possible areas (for odd $N$ this is half of the number of structures tested), and then the product, which is the number of candidates whose perimeter we evaluate. The last five columns give the number of distinct realizable structures found after minimization, for each area ratio. For odd $N$ the candidates with one extra large region are shown in the top row for each $N$.}
\label{tab:numcands}
\end{table}

 \section{Results}

\subsection{Least perimeter candidates at representative area ratios}

The perimeter $P$ decreases quite strongly with increasing area ratio, because small enough regions make only a small perturbation to a structure with lower $N$, and structures with lower $N$ have lower $P$. Although the average area of each region is fixed (at $1/N$), the polydispersity increases with $A_r$. A general measure of polydispersity for regions $i$ with areas $A_N^i$ is 
\begin{equation}
 p = \frac{  \sqrt{\langle A_N^i \rangle} } { \langle \sqrt{A_N^i} \rangle } -1,
\end{equation}
where $\langle \; \rangle$ denotes an average over $i$. Note that with this definition $p=0$ for a monodisperse partition. For a partition with $N_L$ large regions this becomes
\begin{equation}
 p = \frac{\sqrt{\frac{N_L}{N} A_r+ (1-\frac{N_L}{N})}} { \frac{N_L}{N} \sqrt{A_r} + (1-\frac{N_L}{N})} -1.
\end{equation}
We expect the perimeter to decrease as $1/(1+p)$~\cite{kraynikrvs04}, and so to help distinguish different candidates for given $N$ over a range of area ratio $A_r$, we plot $P(1+p)$ in the following.

\begin{figure}
 	\centerline
 	{
 	\includegraphics[width=0.5\linewidth]{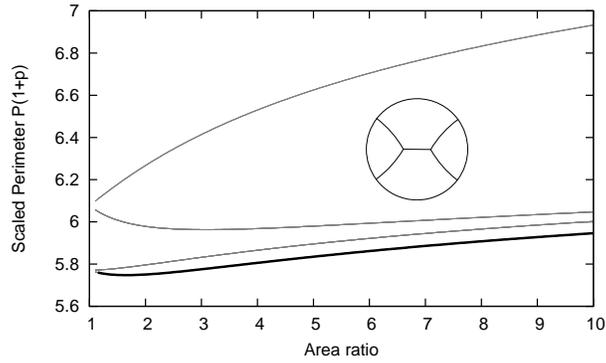}
 	}
 	\caption{The perimeter $P$ of the least perimeter candidates $N=4$ at different area ratios.}
 	\label{fig:energies4}
 \end{figure}

\begin{figure}
 	\centerline
 	{
 	(a)
 	\includegraphics[width=0.5\linewidth]{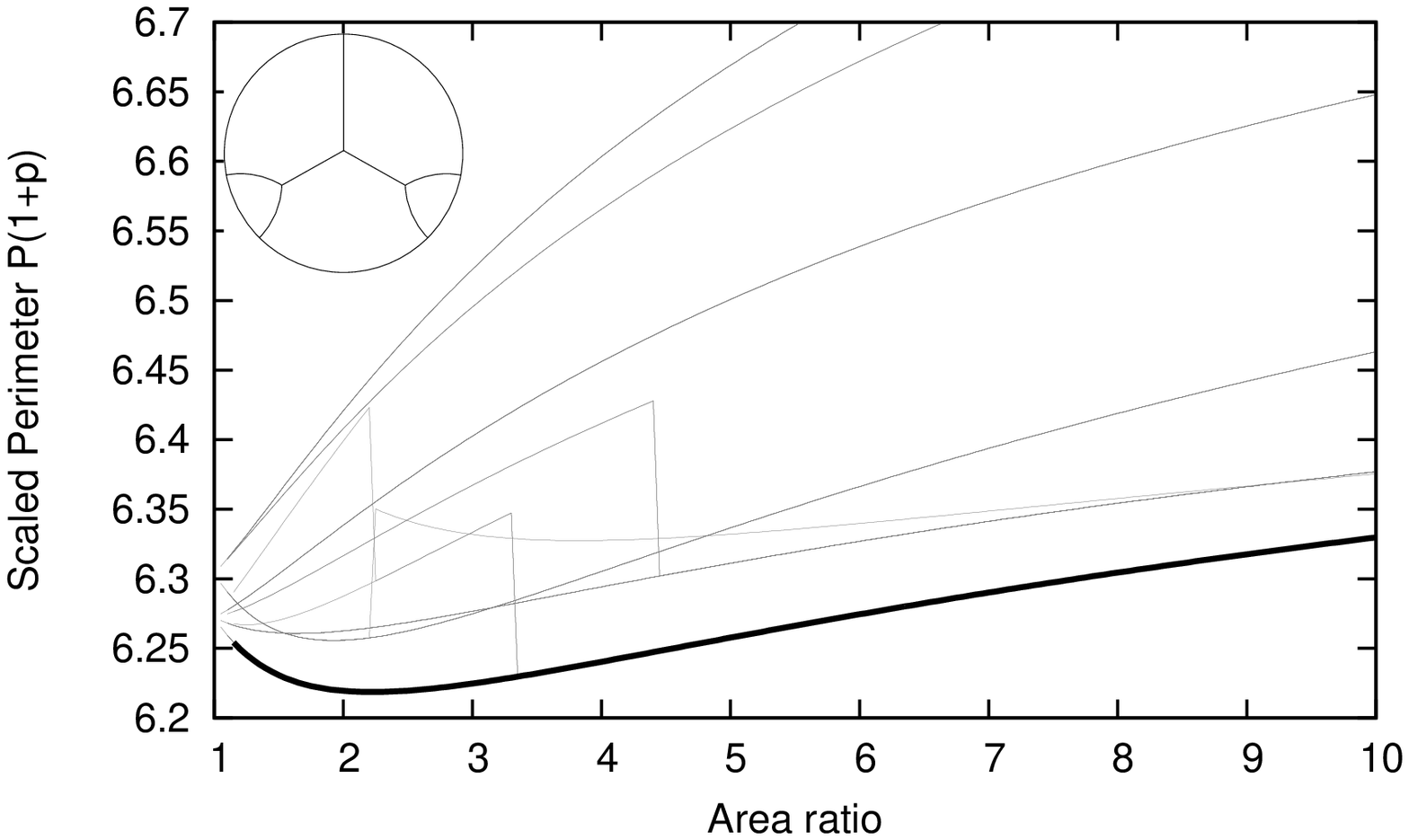}
	}
 	\centerline
 	{
 	(b)
 	\includegraphics[width=0.5\linewidth]{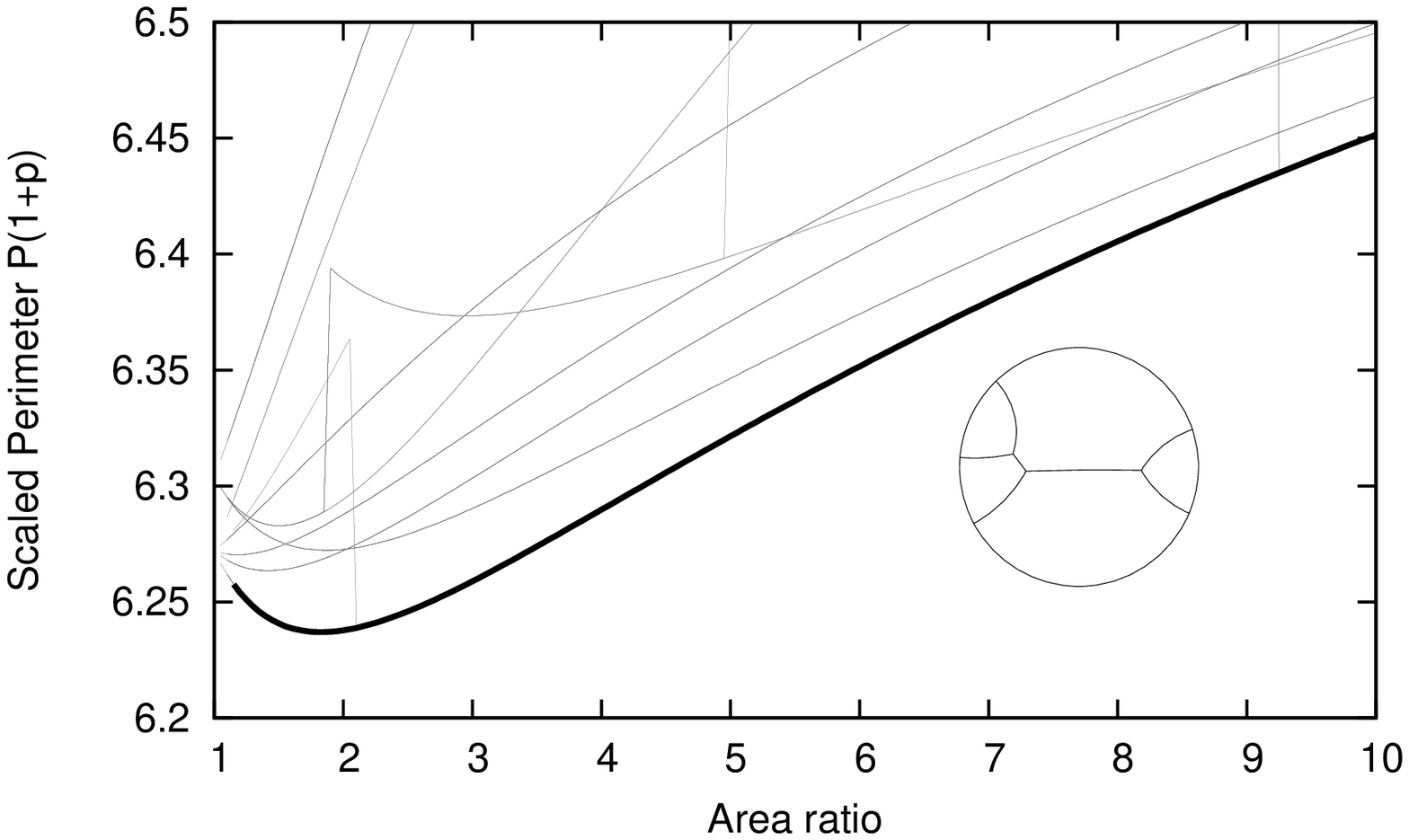}
 	}
 	\caption{The perimeter $P$ of the least perimeter candidates $N=5$ at different area ratios, for (a) the case with one extra large region $5_{32}$ and (b) one extra small region $5_{23}$. 
 	Sudden drops in $P$ correspond to topological changes when an edge shrinks to zero length.
 	}
 	\label{fig:energies5}
 \end{figure}

\begin{figure}
 	\centerline
 	{
 	\includegraphics[width=0.5\linewidth]{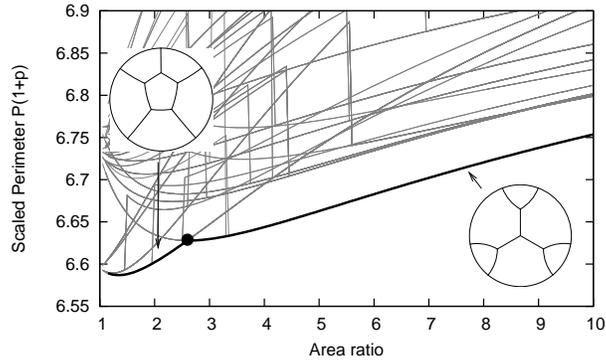}
 	}
 	\caption{The perimeter $P$ of the least perimeter candidates $N=6$ at different area ratios.  
 	The transition between the two optimal structures are marked by a black dot.}
 	\label{fig:energies6}
 \end{figure}

\begin{figure}
 	\centerline
 	{
 	(a)
 	\includegraphics[width=0.5\linewidth]{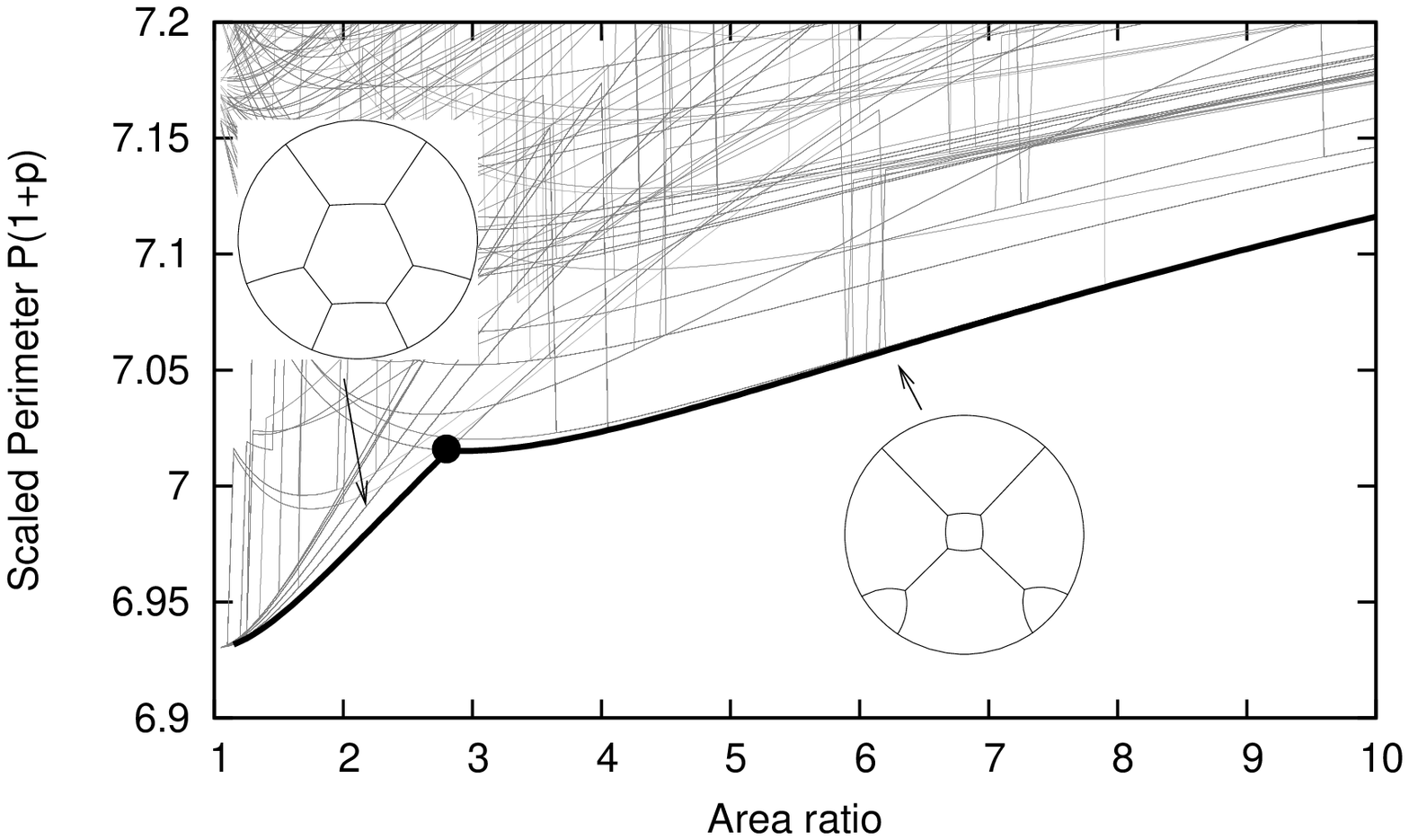}
	}
 	\centerline
 	{
 	(b)
 	\includegraphics[width=0.5\linewidth]{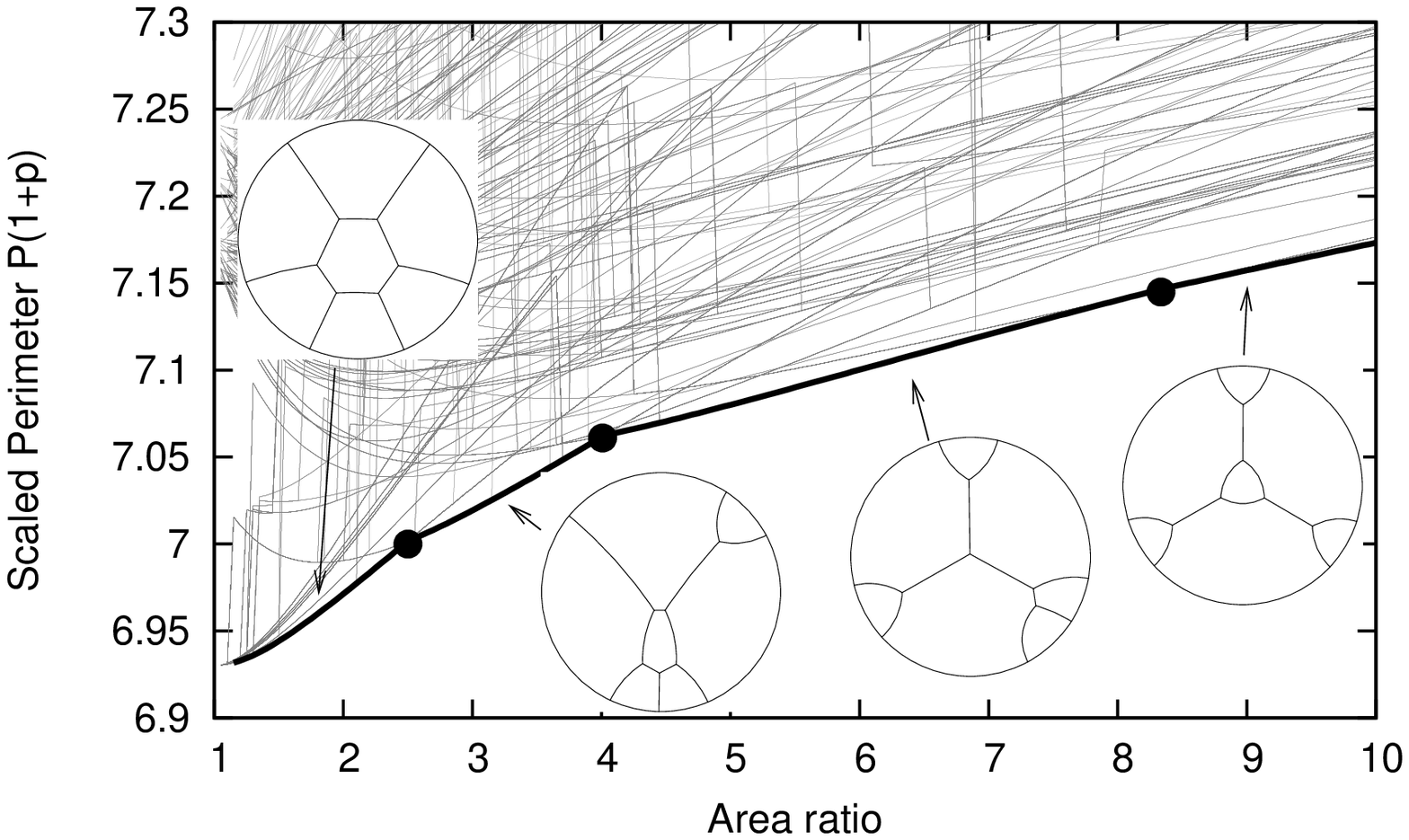}
 	}
 	\caption{The perimeter $P$ of the least perimeter candidates $N=7$ at different area ratios, for (a) the case with one extra large region $7_{43}$ and (b) one extra small region $7_{34}$.
 	}
 	\label{fig:energies7}
 \end{figure}

\begin{figure}
 	\centerline
 	{
 	\includegraphics[width=0.5\linewidth]{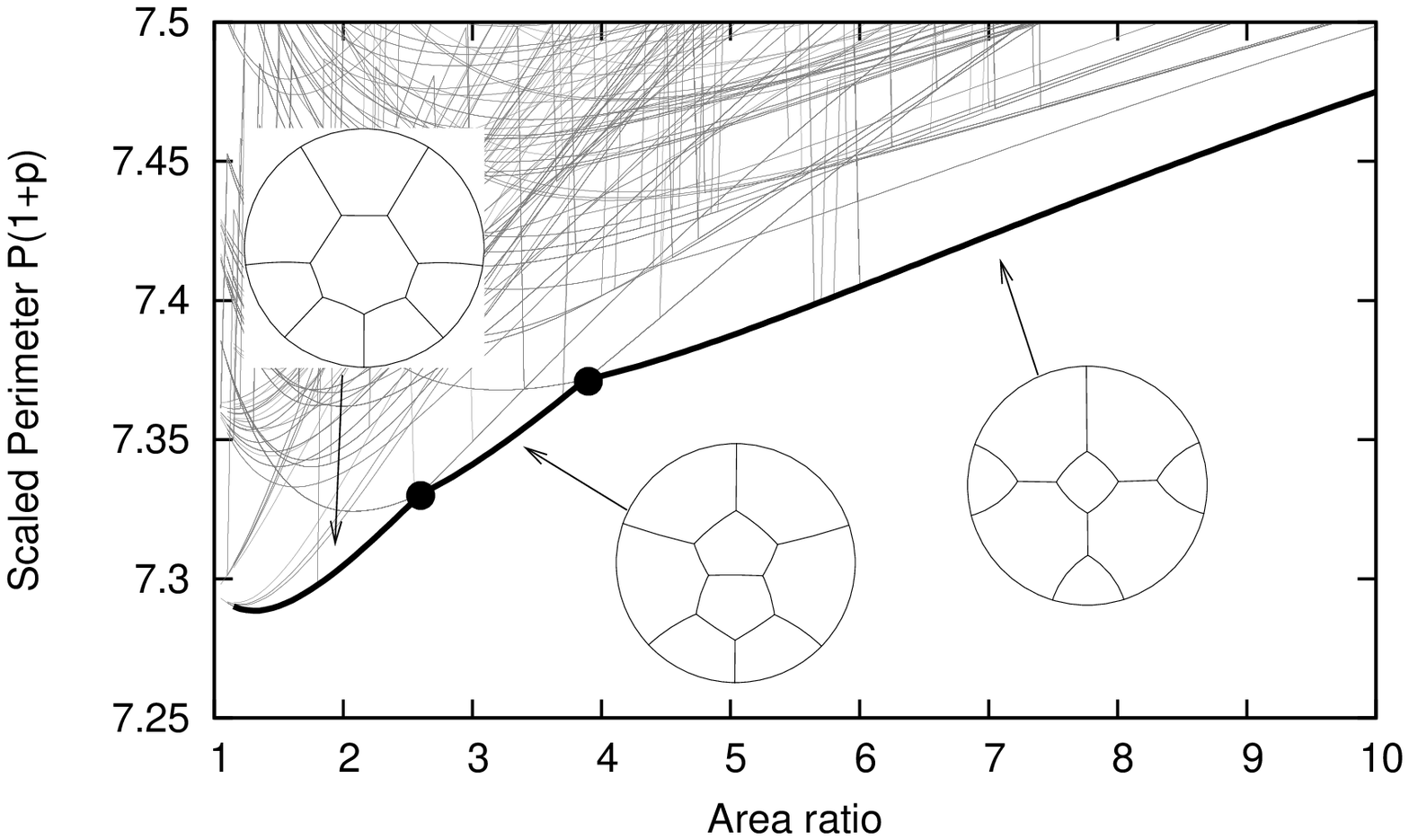}
 	}
 	\caption{The perimeter $P$ of the least perimeter candidates $N=8$ at different area ratios. 
 	We only show the perimeter corresponding to the fifty best candidates for each area ratio.}
 	\label{fig:energies8}
 \end{figure}

\begin{figure}
 	\centerline
 	{
 	(a)
 	\includegraphics[width=0.5\linewidth]{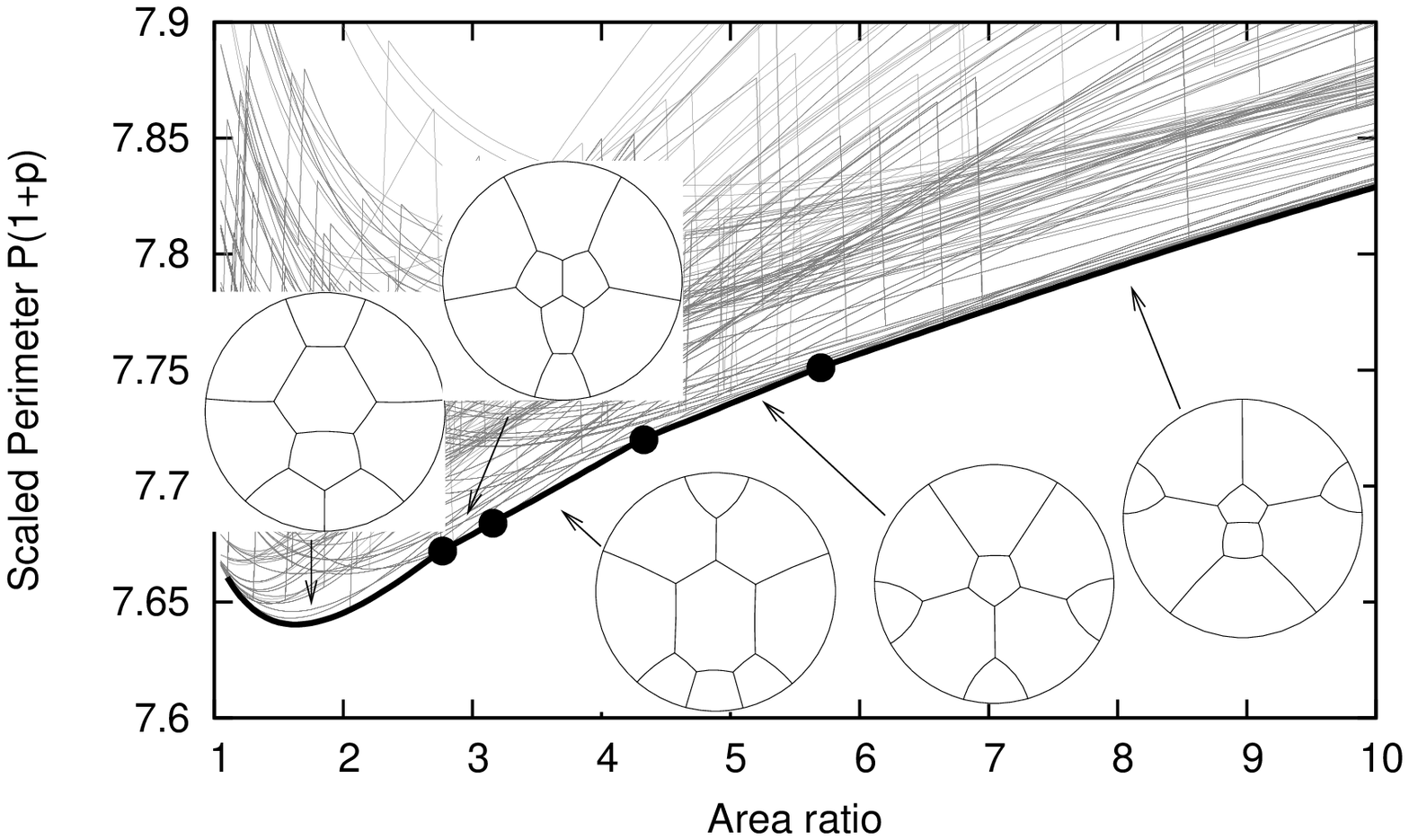}
	}
 	\centerline
 	{
 	(b)
 	\includegraphics[width=0.5\linewidth]{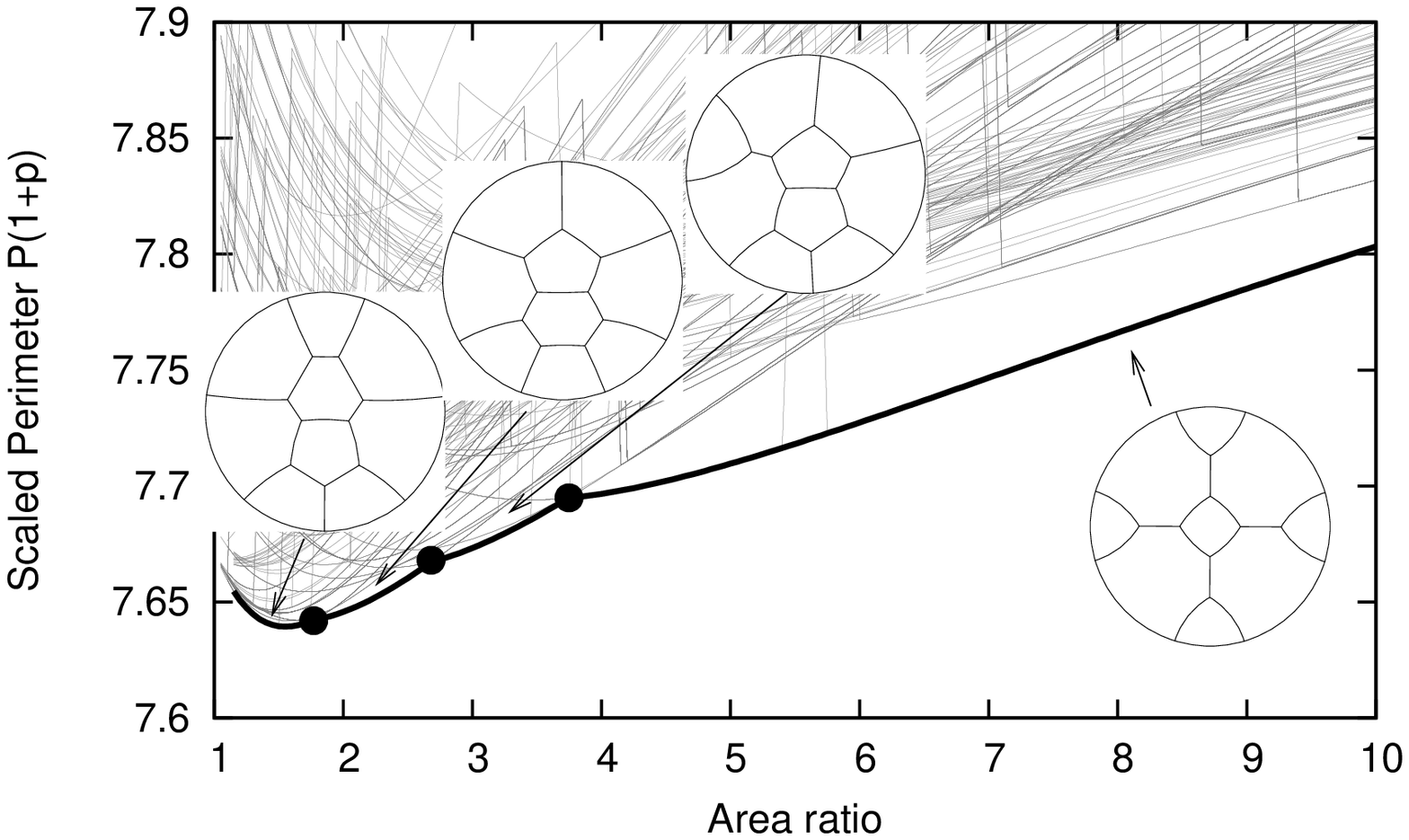}
 	}
 	\caption{The perimeter $P$ of the least perimeter candidates $N=9$ at different area ratios, for (a) the case with one extra large region $9_{54}$ and (b) one extra small region $9_{54}$. 
 	We only show the perimeter corresponding to the fifty best candidates for each area ratio.
 	}
 	\label{fig:energies9}
 \end{figure}

\begin{figure}
 	\centerline
 	{
 	\includegraphics[width=0.5\linewidth]{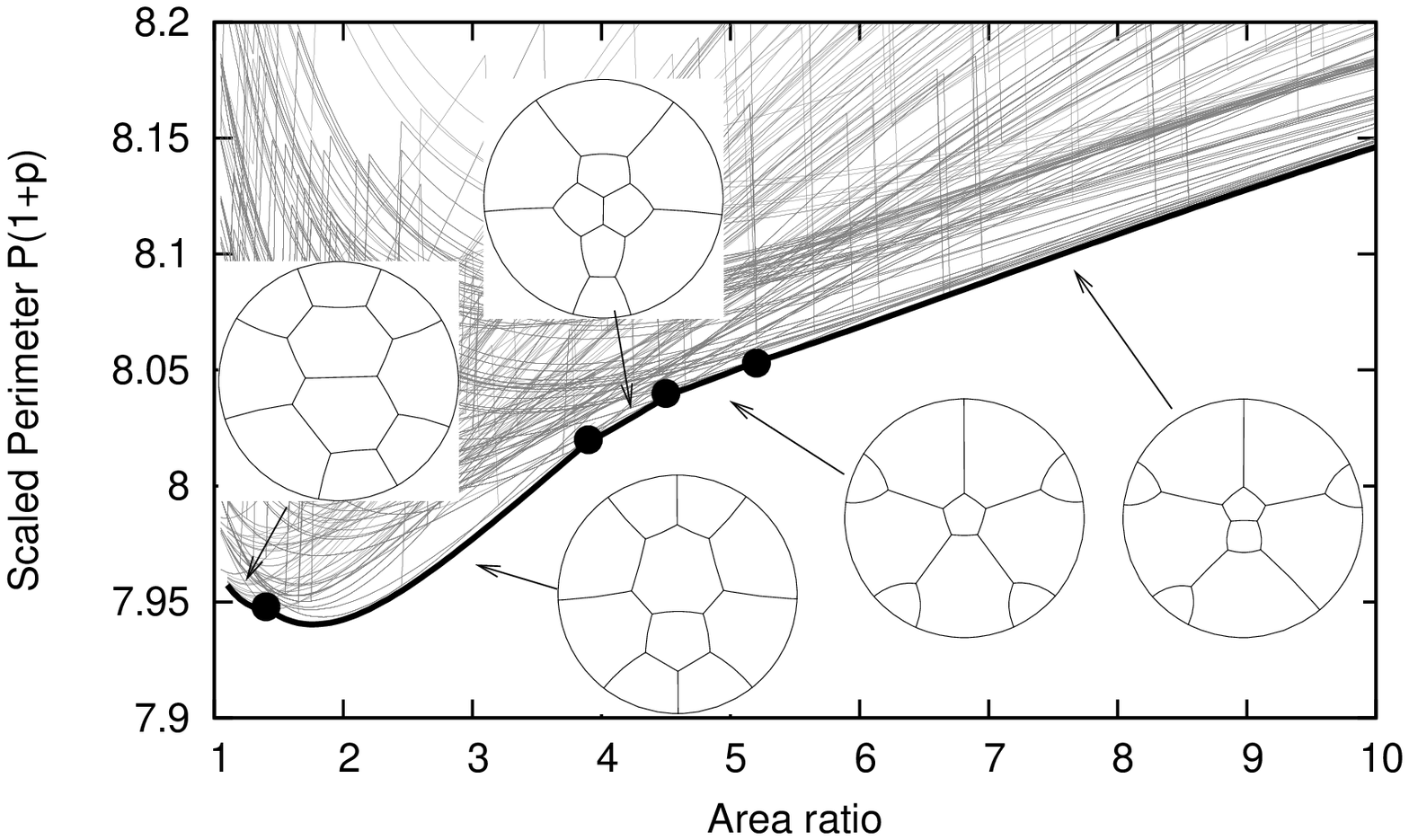}
 	}
 	\caption{The perimeter $P$ of the least perimeter candidates $N=10$ at different area ratios. 
 	We only show the perimeter corresponding to the fifty best candidates for each area ratio.}
 	\label{fig:energies10}
 \end{figure}

Figures~\ref{fig:energies4}--\ref{fig:energies10}, for $N=4$ to $10$ 
respectively, show the scaled perimeter of the structures analysed. The 
optimal perimeter for each $N$ and each $A_r$ is highlighted with a 
thick line, the transitions between structures are indicated, and the 
least perimeter structures themselves are shown according to the area 
ratio at which they are found.

We start by investigating area ratios $A_r = 2, 4, 6, 8$ and $10$. For $N=4$ and $5$ there is no change in the topology of our conjectured least perimeter structure as the area ratio changes; see figures \ref{fig:energies4} and \ref{fig:energies5}. For $N=4$ the two smaller regions never touch, and lie at opposite ends of a straight central edge. For $N=5$, for both possible distributions of large and small regions, the optimal pattern always consists of two three-sided regions whose internal vertices are connected to the other internal vertex, which itself has one other connection to the boundary of the disc. That is, in neither case does the optimal candidate have an internal region.

For $N \geq 6$ there are transitions between different structures as the area ratio $A_r$ changes. We therefore interpolate between these values of area ratio to determine the critical values of $A_r$ at which the changes in topology of the least perimeter candidate occur for each $N$.

We do this by taking each of the structures that was found for each area ratio $A_r = 2, 4, \ldots$ and change the area ratio in small steps (of 0.05). For each of these candidates we find and record the perimeter. (For $N=10$ we do this only for the fifty or so best candidates for each value of 
$A_r$, since there are so many candidates which are far from optimal for any area ratio.) 
For candidates whose initial area ratio was 2, 4 or 6 we decreased the area ratio to 1.1 and the increased it up to 10. For candidates whose initial area ratio was 8 or 10 we increased the area 
ratio up to 10 before slowly decreasing it down to 1.1. We are therefore able to confirm that at low enough area ratio we recover the optimal structures found in the monodisperse case~\cite{Cox06}. 

This procedure generates a few extra optimal structures that are missed by the first sampling of the area ratios, for example between $A_r = 2$ and 4 for $N=8, 9_{54}$ and $9_{45}$ and between $A_r = 4$ and $6$ for $N=9_{54}$ and for $N=10$.

For $N=6$ we find that the topology of the least perimeter candidates 
with area ratios $A_r \ge 2.6$ are the same (figure~\ref{fig:energies6}). For area ratios less than this value the topology is that of the optimal candidate in the monodisperse case~\cite{Cox06}.

The two different cases for $N=7$ behave differently (figure~\ref{fig:energies7}). In the case $7_{43}$ there are only two different candidates found, and for $A_r \ge 2.8$ the topology does not change. On the other hand, for $7_{34}$ we find four different topologies, with a transition to a new candidate at a surprisingly high area ratio of 8.4.

The least perimeter structure with $N=8$ regions has the monodisperse topology for $A_r < 2.6$ (figure~\ref{fig:energies8}); there is one further transition at $A_r = 3.9$, giving three different optimal structures.

For $N=9$ the results are richer (figure~\ref{fig:energies9}), in the sense that the system explores more possible states as the area ratio changes. For $9_{54}$ we find five different topologies, while for $9_{45}$ there are four. In the latter case the structure found for $A_r = 2$ is different to the monodisperse one~\cite{Cox06}, and there is a transition to that structure at a low area ratio around 1.8.

Finally, for $N=10$ (figure~\ref{fig:energies10}) we again find a candidate for $A_r = 2$ that differs from the monodisperse case and a transition at even lower area ratio. In total there are five different topologies.

\subsection{Analysis of patterns}

\begin{figure}
 	\centerline
 	{
 	\includegraphics[width=0.5\linewidth]{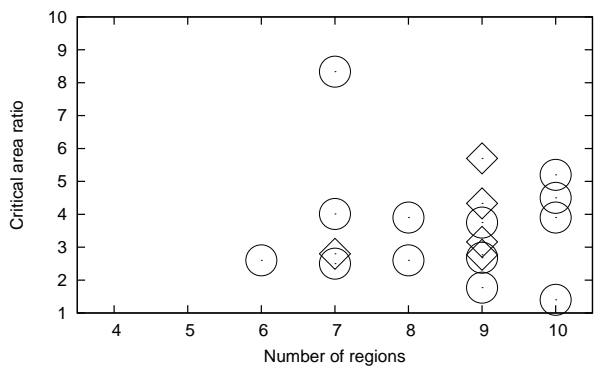}
 	}
 	\caption{
 	The area ratio at which there is a transition between different least perimeter arrangements.
 	 For odd $N$, diamonds refer to the case with one extra large region.
 	 }
 	\label{fig:crossover}
 \end{figure}

The critical area ratios at which there is a transition between optimal 
structures are summarised in figure \ref{fig:crossover}. Most are found 
at intermediate values of the area ratio, roughly between $A_r = 2.5$ 
and 4, although this broadens slightly with increasing $N$. There is 
also a single point at high $A_r$, for $N=7$, which corresponds to 
moving a small bubble from the boundary of the disc to the centre, and 
hence to a symmetric state. It is perhaps surprising that this highly symmetric 
state is not optimal at lower area ratio, since many of the least 
perimeter structures {\em are} symmetric.

\begin{figure}
 	\center
 	\includegraphics[width=0.5\linewidth]{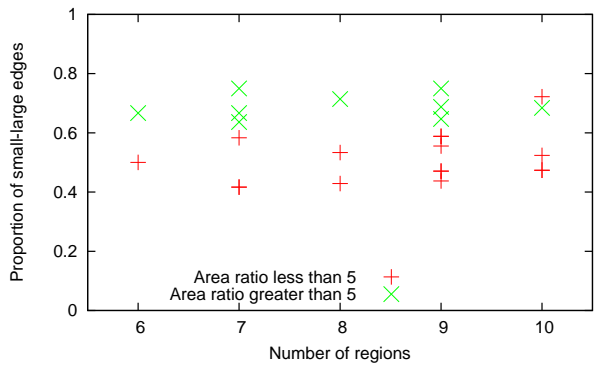}
 	\caption{The proportion of edges separating large from small regions $E_{ls}$ in the least perimeter candidate for each $N$. The area ratio is distinguished by whether it is greater or less than five. Data for the cases of an extra large or an extra small region is collated.} 
 	\label{fig:largesmall}
 \end{figure}

The images in figures~\ref{fig:energies4}--\ref{fig:energies10} also hint at an evolution from the small regions clustering together at low area ratio to being separated from each other by the large regions at high area ratio.
We quantify this observation by counting the proportion of edges $E_{LS}$ separating large from small regions in each least perimeter structure. A structure with a higher value of $E_{LS}$ has less clustering. The data in figure \ref{fig:largesmall} bears out this observation: for $N \ge 6$ and small area ratio the value of $E_{LS}$ is lower than for large area ratio. (The exception is one of the structures for $N=10$, where even at low area ratio ($4.5 \le A_r \le 5.2$) the least perimeter candidate has the small bubbles well-separated.)

\section{Conclusions}

We have enumerated all candidate partitions of the disc with $N \le 10$ 
regions with one of two different areas, and determined, for each area 
ratio, the partition with least perimeter. The results show an 
increasing number of transitions between the different optimal 
structures found for varying area ratio as $N$ increases, mostly at low 
area ratio. Further, in the least perimeter partitions at small area 
ratio the smaller regions are clustered together, while at large area 
ratio the small regions are separated by large regions. Transitions between such mixed and sorted configurations often occur as a consequence of some agitation~\cite{vazct11}.

The procedure described here should translate directly to least 
perimeter partitions of the surface of a sphere, since to enumerate 
candidates to that problem we are able to use the same graphs and 
consider the periphery of the graph to form the boundary of one further 
region. Thus the candidates for the disc with $N$ regions are also the 
candidates for the sphere with $N+1$ regions. In general, our 
preliminary results indicate, as for the monodisperse 
case~\cite{Coxf10}, that the least perimeter arrangement of regions on 
the sphere is {\em different} to the corresponding optimal partition of 
the disc.

\section*{Acknowledgements}

We thank DG Evans for helpful discussions. FJH was supported by a Walter 
Idris Jones Research Scholarship and SJC by the UK Engineering and 
Physical Sciences Research Council (EP/N002326/1).

\end{document}